\documentclass[final,3p,12pt]{elsarticle}

\usepackage{graphicx,subfigure}

\usepackage{amssymb,amsmath}

\usepackage[colorlinks,linkcolor=blue,citecolor=blue,urlcolor=blue]{hyperref}

\biboptions{compress}

\journal{Optics Communications}

\begin{document}

\def\<{\langle}
\def\>{\rangle}

\graphicspath{{figures/}}

\begin{frontmatter}



\title{Two-Photon Scattering by a Cavity-Coupled Two-Level Emitter in a One-Dimensional Waveguide}


\author{Zhongling Ji and Shaoyan Gao\corref{cor1}}

\cortext[cor1]{Corresponding Author: gaosy@mail.xjtu.edu.cn}

\address{Department of Applied Physics, MOE Key Laboratory for Nonequilibrium Synthesis and Modulation of Condensed Matter, Xi'an Jiaotong University, Xi'an 710049, China}

\begin{abstract}
We show that two-photon transport can be modulated by a two-level emitter coupled to a cavity in a one-dimensional waveguide. In the ordinary case, the transmitted light has a wider frequency spectrum than the situation without the cavity because it is reflected and scattered many times. But when the two photons are resonant with the cavity resonance reflection frequency, the frequency spectrum of the transmitted light becomes narrower than that without the cavity. This means that properly tuning the cavity resonance frequency can improve the photon-photon interaction. In addition, we show that the two-photon intensity correlation functions are nearly opposite to each other at the two sides of the emitter transition frequency rather than be the same, which is exactly the Fano resonance line shape for two photons. Such an effect is important for lowering the power threshold in optical bistable devices and for sensing applications. When the emitter transition frequency equals to the cavity resonance frequency for a high-Q cavity, our results agree with the recent experiments and theories.
\end{abstract}


\end{frontmatter}


\section{Introduction}\label{sec:1}
Quantum internet is important for quantum computation, communication and metrology \cite{KimbleNat453}. The main methods to realize the quantum networks are the cavity-QED-based protocol \cite{CiracPRL78} and the protocol provided by Duan, Lukin, Cirac, and Zoller (DLCZ) \cite{DuanNat414, ChoiNat452}. The former one uses the interaction between light and $\Lambda$-type three-level atoms in a high-Q cavity to store quantum states \cite{FleischOC179}. The latter one involves the measurement-induced entanglement \cite{CabrilloPRA59}, replaces single atoms by atomic ensembles so that it reduces the restriction of the cavity's quality and utilize the built-in entanglement-purification to create the long-distance quantum communication efficiently.

Thus it is of great significance to consider the interaction between few photons and a cavity-coupled emitter. Here we focus on a unique one-dimensional (1D) waveguide for photons because it is helpful to realize the strong photon-photon interaction \cite{FanPRL98, FanPRA76}. Recently, people have already done some researches about the interaction between light and an atom-cavity system in a 1D waveguide both in experiment \cite{WallraffNat431, BirnbaumNat436} and in theory \cite{FanAPL80, FanJOSAA20, FanOL30, FanPRL95, BlaisPRA69, FanPRA79, ShiarXiv1009}. These theoretical approaches include either scattering matrix \cite{FanAPL80, FanJOSAA20, FanOL30, FanPRL95} or quantum field theory \cite{BlaisPRA69, FanPRA79, ShiarXiv1009}. The two methods get the similar results for single photons \cite{FanPRL95, BlaisPRA69}.

In this paper we study the interaction of two photons and a cavity-coupled two-level emitter in a 1D waveguide. By analyzing two-photon wave packets, generalizing the scattering matrix \cite{HausWFO} to dispose frequency-spectrum transformation, we get the two-photon correlated momenta distribution, correlated transmitted coefficient, Fano resonance line shape and other important properties. Comparing with most existing works, our results are more general because our approach is not restricted to a high-Q cavity. Because our model includes the decay inherently, our results exactly agree with the experimental results in Ref. \cite{BirnbaumNat436}. In Sec.~\ref{sec:2} we show the model and analyze two-photon overlaps. In Sec.~\ref{sec:3} we generalize the scattering matrix \cite{HausWFO} to deal with the frequency-spectrum transformation. We show the results in Sec.~\ref{sec:4} and make a summary in Sec.~\ref{sec:5}.

\section{The Model and Two-Photon Wave Packets}\label{sec:2}
\begin{figure}
  \centering
  \subfigure[Experimental setup]{\label{fig:1a}
  \includegraphics[height=1in]{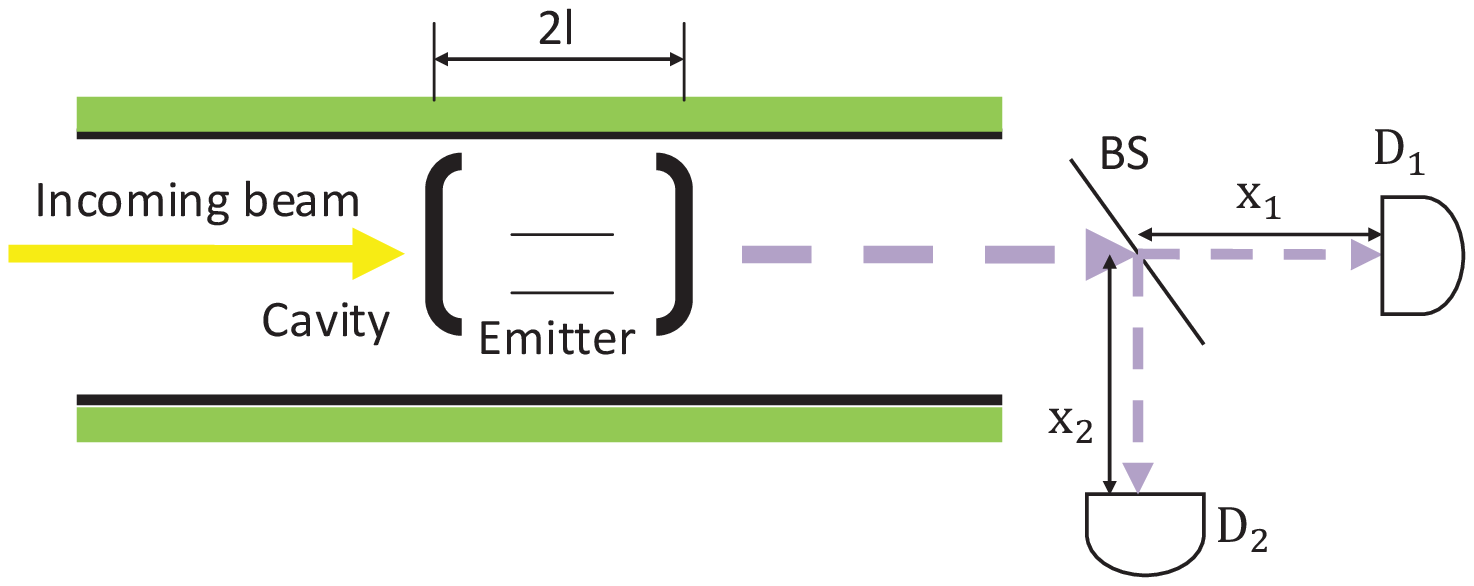}}\hspace{0.5in}
  \subfigure[Two-level emitter]{\label{fig:1b}
  \includegraphics[height=1in]{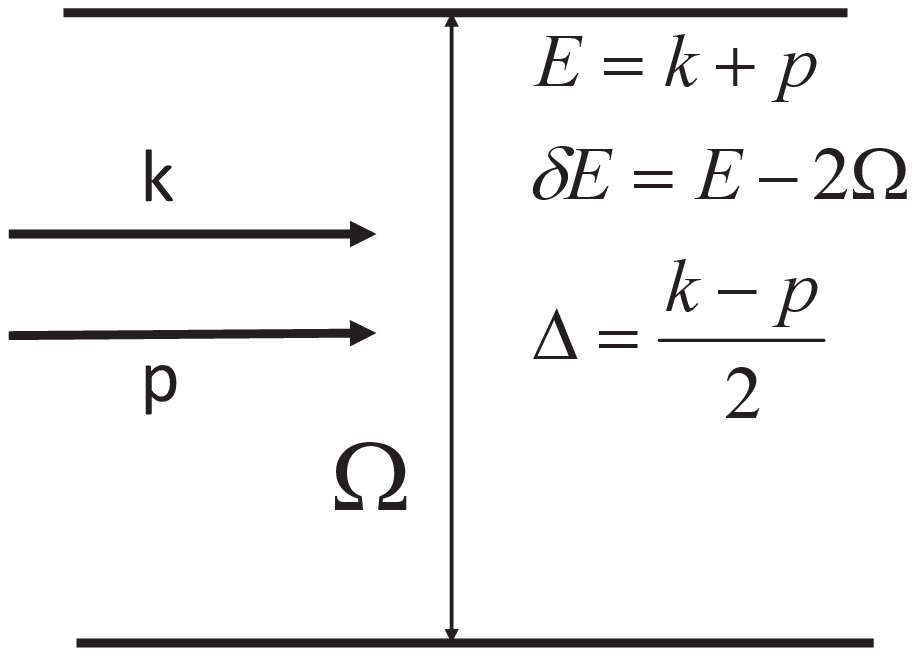}}\\
  \caption{(a) The schematic experimental setups for concurrence measurement of the correlated transmission probability density $|t_2(x_1,x_2)|^2$. $D_1$, $D_2$ are photon-detectors with adjustable positions. BS is beam splitter. The ``='' denotes the two-level emitter in the 1D waveguide. The black parentheses symbolize the cavity, of which the length is $2l$. (b) The two-level emitter. $k$ and $p$ are the momenta of the two photons. $\Omega$ is the emitter transition frequency.}\label{fig:1}
\end{figure}

There are several methods to realize the atom-cavity system in a 1D waveguide \cite{WallraffNat431, BirnbaumNat436, FanAPL80}. Fig.~\ref{fig:1a} shows the schematic experimental setups \cite{FanPRA76}. The incident beam is monochrome plane wave. After the two photons are scattered by the whole system, we separate them by a beam splitter (BS) and use two single-photon counters to detect them. We register the probability of concurrent detections. Similarly with the Hanbury Brown-Twiss effect \cite{ScullyQO}, the results can show correlation of the two photons. Fig.~\ref{fig:1b} shows the parameters of the two-level emitter.

\begin{figure}
  \centering
  \subfigure[Two-photon wave packets]{\label{fig:2}
  \includegraphics[height=2.5in]{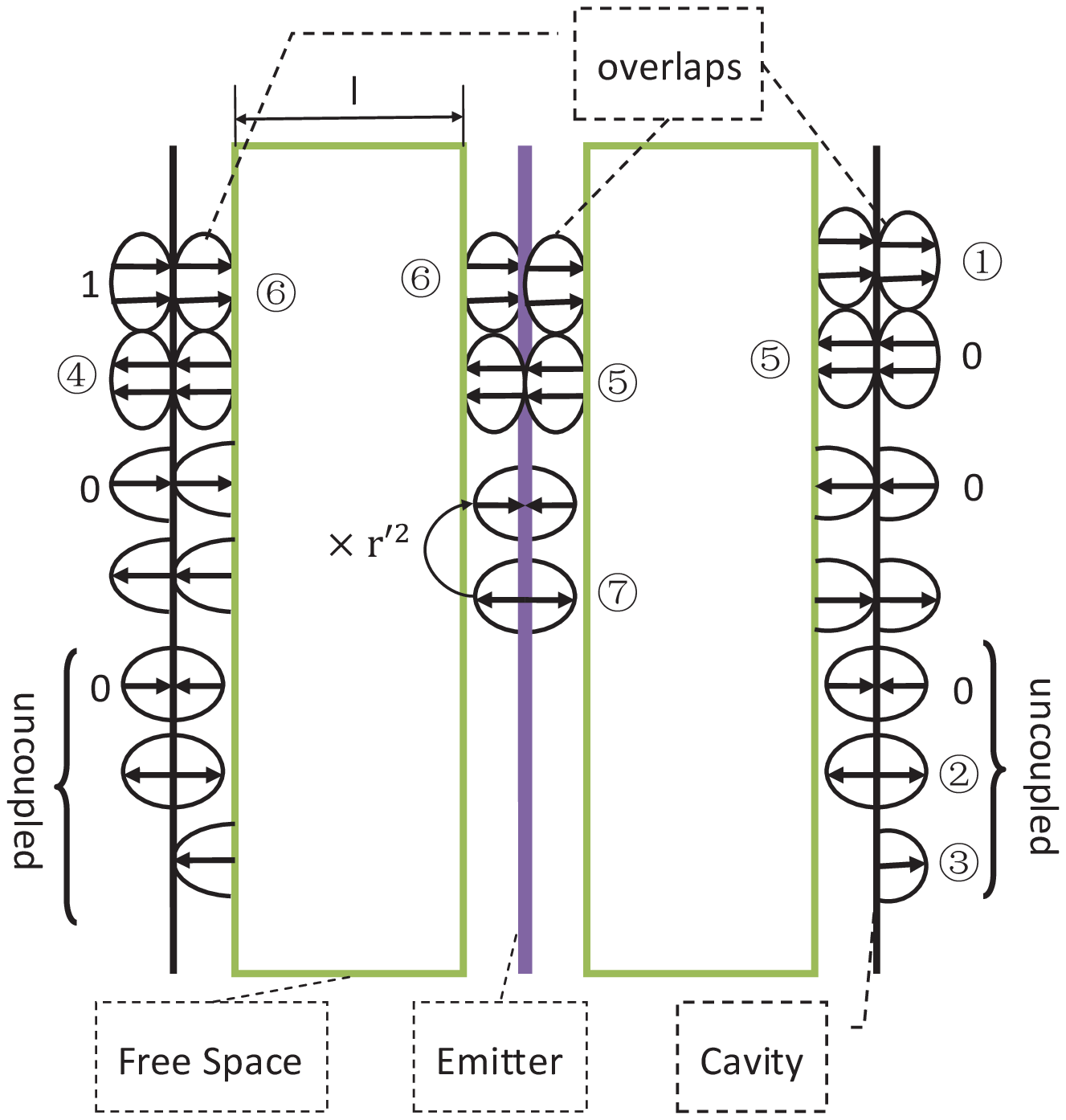}}\hspace{0.5in}
  \subfigure[Symbols in Eq.~(\ref{eq:1})]{\label{fig:2b}
  \includegraphics[height=2.5in]{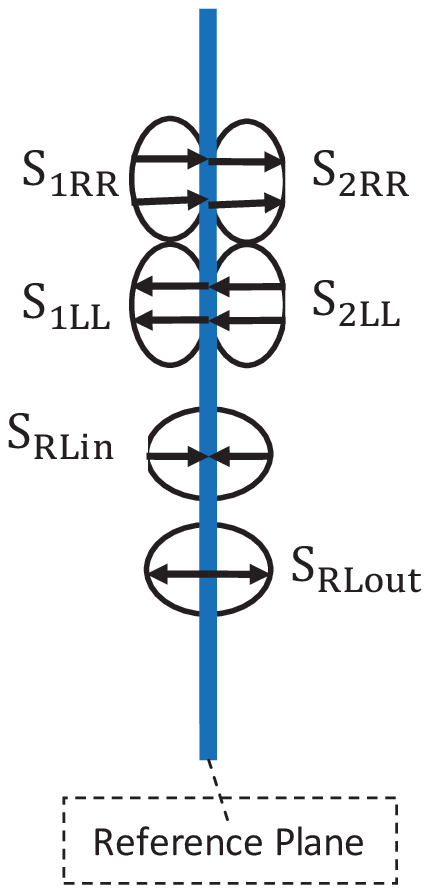}}\\
  \caption{(a) The ellipses are the two-photon wave packets, of which some are at together and some are separated. The arrows in the ellipses represent the moving directions of the photons. The uncircled numbers, e.g. ``1'' and ``0'', denote the known amplitudes of wave packets each. The black, green and purple lines are the reference plane of the cavity, the free space and the emitter respectively. (b) Symbols in Eq.~(\ref{eq:1}). $S_{1RR}$ et al. are frequency spectrums of the wave packets. The blue line is the reference plane.}\label{fig:2:0}
\end{figure}

Fig.~\ref{fig:2} shows wave packets of the two photons. After the incidence of the two photons on the reference plane, they can be scattered to different sides of the plane. Incident photons can also come from different sides of the system. So we cannot divide these wave packets into left-coming or right-coming parts. However, we find that if each one of the two photons runs out of the cavity, it cannot form the situation that two photons are at together again. So we take this situation as ``loss'' at first [see the ``uncoupled'' wave packets in Fig.~\ref{fig:2}]. After calculation of wave packet \textcircled{\small{1}}, we add wave packets \textcircled{\small{2}} and \textcircled{\small{3}} to make the complete resolution of this problem.

\section{Scattering Matrix and Frequency-Spectrum Transformation}\label{sec:3}
For two-photon case, the emitted light has a frequency-spectrum distribution \cite{FanPRL98, FanPRA76}. So we need to generalize the scattering matrix \cite{HausWFO} to dispose this transformation. In fact, we need to solve several integral equations to get the spectrums of the outgoing wave packets [wave packets \textcircled{\small{1}}, \textcircled{\small{2}} and \textcircled{\small{3}} in Fig.~\ref{fig:2}] from the incident wave packets [the wave packets denoted by uncircled numbers ``1'' and ``0'' in Fig.~\ref{fig:2}], of which the amplitudes are known. Here we use scattering matrix to express and solve these equations. We define the scattering matrix as follow
\begin{equation}\label{eq:1}
    \begin{bmatrix}
      S_{1LL}(k,p) \\
      S_{2RR}(k,p) \\
      S_{RLout}(k,p) \\
    \end{bmatrix}
    =
    \begin{bmatrix}
      \hat{R}_{22} & \hat{T}_{22} & \hat{T}_{12} \\
      \hat{T}_{22} & \hat{R}_{22} & \hat{T}_{12} \\
      \hat{T}_{21} & \hat{T}_{21} & \hat{T}_{11} \\
    \end{bmatrix}
    \begin{bmatrix}
      S_{1RR}(k,p) \\
      S_{2LL}(k,p) \\
      S_{RLin}(k,p) \\
    \end{bmatrix}.
\end{equation}
Here, $S_{1RR}(k,p)$ et al. are frequency spectrums of the wave packets [Fig.~\ref{fig:2b}]. The operator $\hat{T}_{22}$ ($\hat{R}_{22}$) represents that two photons come together and transmit (reflect) together. $\hat{T}_{12}$ describes that two photons come from different sides of the reference plane and are scattered to the same side, of which the reverse process is denoted by $\hat{T}_{21}$. $\hat{T}_{11}$ denotes that two photons come from and to different sides of the system. According to \cite{FanPRA76} we have
\begin{equation}\label{eq:2}
\begin{gathered}
    \hat{T}_{22} f(k,p) = \frac{1}{2} \int_{0}^{\infty} dk_1 dp_1 f(k_1,p_1)\times \\
    \{ \bar{t}_{k_{1}} \bar{t}_{p_{1}} [ \delta(k_{1}-k) \delta(p_{1}-p) + \delta(k_{1}-p) \delta(p_{1}-k) ] + B \delta(E_{1}-E) \},
\end{gathered}
\end{equation}
\begin{equation}\label{eq:3}
\begin{gathered}
    \hat{R}_{22} f(k,p) = \frac{1}{2} \int_{0}^{\infty} dk_1 dp_1 f(k_1,p_1)\times \\
    \{ \bar{r}_{k_{1}} \bar{r}_{p_{1}} [ \delta(k_{1}-k) \delta(p_{1}-p) + \delta(k_{1}-p) \delta(p_{1}-k) ] + B \delta(E_{1}-E) \},
\end{gathered}
\end{equation}
\begin{equation}\label{eq:4}
\begin{gathered}
    \hat{T}_{12} f(k,p) = \frac{1}{\sqrt{2}} \hat{T}_{21} f(k,p) = \frac{1}{2} \int_{0}^{\infty} dk_1 dp_1 f(k_1,p_1)\times \\
    \left\{ \frac{\bar{t}_{k_{1}} \bar{r}_{p_{1}} + \bar{r}_{k_{1}} \bar{t}_{p_{1}}}{2} [ \delta(k_{1}-k) \delta(p_{1}-p) + \delta(k_{1}-p) \delta(p_{1}-k) ] + B \delta(E_{1}-E) \right\},
\end{gathered}
\end{equation}
\begin{equation}\label{eq:5}
\begin{gathered}
    \hat{T}_{11} f(k,p) = \frac{1}{2} \int_{0}^{\infty} dk_1 dp_1 f(k_1,p_1)\times \\
    \left\{ \frac{\bar{t}_{k_{1}} \bar{t}_{p_{1}} + \bar{r}_{k_{1}} \bar{r}_{p_{1}}}{\sqrt{2}} [ \delta(k_{1}-k) \delta(p_{1}-p) + \delta(k_{1}-p) \delta(p_{1}-k) ] + \sqrt{2}B \delta(E_{1}-E) \right\},
\end{gathered}
\end{equation}
where $f(k,p)=f(p,k)$ denotes the frequency spectrum of the incident photons. $\bar{t}_{k}$ and $\bar{r}_{k}$ are the transmission and reflection amplitude of the emitter for single photons respectively \cite{FanPRA76}. $B$ is the background fluorescence
\begin{equation}\label{eq:6}
    B = \frac {4i \Gamma^2} {\pi} \frac {E_1 - 2 \Omega + i\Gamma} {[4\Delta_1^2 - (E_1 - 2\Omega + i\Gamma)^2] [4\Delta^2 - (E_1 - 2\Omega + i\Gamma)^2]}.
\end{equation}
Here, $\Gamma$ is the coupling strength between the photons and the emitter.

From Eq.~(\ref{eq:2}) we know that operator $\hat{T}_{22}$ has two parts: the one is multiplied by $[\delta(k_{1}-k) \delta(p_{1}-p) + \delta(k_{1}-p) \delta(p_{1}-k)]$ and the other is multiplied by $\delta(E_{1}-E)$. The calculation of operators can be considered by each part respectively. In Tab.~\ref{tab:1} we have the operator-multiplying formula (see \ref{app:1} for details)
\begin{table}[h]
\centering
\caption{Operator-multiplying formula}\label{tab:1}
\begin{tabular}{ccc}
    \hline
    Operators & $\delta(E_{1}-E)$ & $\delta(k_{1}-k)\delta(p_{1}-p)$ \\ \hline
    $\hat{T}$ & $B_0B_1(\Delta_1)B_2(\Delta_2)$ & $tt(\Delta)$ \\
    $\hat{R}$ & $C_0C_1(\Delta_1)C_2(\Delta_2)$ & $rr(\Delta)$ \\
    $\hat{R}\hat{T}$ & $B_0B_1(\Delta_1)B_2(\Delta_2)rr(\Delta_2)$ & $tt(\Delta)rr(\Delta)$ \\
    & + $C_0C_1(\Delta_1)C_2(\Delta_2)tt(\Delta_1)$ & \\
    & + $B_0C_0B_1(\Delta_1)C_2(\Delta_2)f_{int}$ & \\
    \hline
\end{tabular}
\end{table}
\\ where
\begin{equation}\label{eq:6:1}
    f_{int} = \frac{1}{2} \int_{-\infty}^{+\infty} B_2(y) C_1(y) dy.
\end{equation}

For calculation, we need the inverse operator, which satisfies
\begin{equation}\label{eq:7}
\begin{gathered}
    \hat{T}^{-1} \hat{T} f(k,p) = \hat{T} \hat{T}^{-1} f(k,p) = f(k,p) = \frac{1}{2} [f(k,p)+f(p,k)] \\
    = \frac{1}{2} \int dk_1 dp_1 f(k_1,p_1) [ \delta(k_{1}-k) \delta(p_{1}-p) + \delta(k_{1}-p) \delta(p_{1}-k) ].
\end{gathered}
\end{equation}
Then we derive Tab.~\ref{tab:2}, which shows the operator-inversion formula (see \ref{app:2} for details)
\begin{table}[h]
\centering
\caption{Operator-inversion formula}\label{tab:2}
\begin{tabular}{ccc}
    \hline
    Operators & $\delta(E_{1}-E)$ & $\delta(k_{1}-k)\delta(p_{1}-p)$ \\ \hline
    $\hat{T}$ & $B_0B_1(\Delta_1)B_2(\Delta_2)$ & $tt(\Delta)$ \\
    $\hat{T}^{-1}$ & $-\frac{B_0}{1+B_0g_{int}} \frac{B_1(\Delta_1)}{tt(\Delta_1)} \frac{B_2(\Delta_2)}{tt(\Delta_2)}$ & $\frac{1}{tt(\Delta)}$ \\
    \hline
\end{tabular}
\end{table}
\\ where
\begin{equation}\label{eq:8}
    g_{int} = \frac{1}{2} \int_{-\infty}^{+\infty} \frac{B_1(y) B_2(y)}{tt(y)} dy.
\end{equation}

Having the formulae in Tab.~\ref{tab:1} and \ref{tab:2}, we can put the reduced scattering matrix of the emitter as follow (see \ref{app:3} for details)
\begin{equation}\label{eq:9}
    \begin{bmatrix}
      S_{1LL}(k,p) \\
      S_{2RR}(k,p) \\
    \end{bmatrix}
    =
    \begin{bmatrix}
      \hat{R} & \hat{T} \\
      \hat{T} & \hat{R} \\
    \end{bmatrix}
    \begin{bmatrix}
      S_{1RR}(k,p) \\
      S_{2LL}(k,p) \\
    \end{bmatrix},
\end{equation}
where
\begin{equation}\label{eq:10}
\begin{aligned}
    \hat{T} & = \hat{T}_{22} + r'^2 \hat{T}_{12} (1-r'^2\hat{T}_{11})^{-1} \hat{T}_{21}, \\
    \hat{R} & = \hat{R}_{22} + r'^2 \hat{T}_{12} (1-r'^2\hat{T}_{11})^{-1} \hat{T}_{21}.
\end{aligned}
\end{equation}
Here we make the substitution
\begin{equation}\label{eq:11}
    r'^2 = \sqrt{2}e^{2iEl} \times r^2,
\end{equation}
where $r$ is the reflectivity of the cavity, $2l$ is the length of the cavity.

Similarly with the approach for single photons in Ref.~\cite{FanAPL80}, we rewrite Eq.~(\ref{eq:9}) as follow
\begin{equation}\label{eq:11:0}
    \begin{bmatrix}
      S_{2RR}(k,p) \\
      S_{2LL}(k,p) \\
    \end{bmatrix}
    =
    \begin{bmatrix}
      \hat{T}-\hat{R}\hat{T}^{-1}\hat{R} & \hat{R}\hat{T}^{-1} \\
      -\hat{T}^{-1}\hat{R} & \hat{T}^{-1} \\
    \end{bmatrix}
    \begin{bmatrix}
      S_{1RR}(k,p) \\
      S_{1LL}(k,p) \\
    \end{bmatrix}.
\end{equation}

The scattering matrix of the cavity and the free space are
\begin{equation}\label{eq:11:1}
    \begin{bmatrix}
      S_{2RR}(k,p) \\
      S_{2LL}(k,p) \\
    \end{bmatrix}
    =
    \begin{bmatrix}
      t^2-\frac{r^4}{t^2} & \frac{r^2}{t^2} \\
      -\frac{r^2}{t^2} & \frac{1}{t^2} \\
    \end{bmatrix}
    \begin{bmatrix}
      S_{1RR}(k,p) \\
      S_{1LL}(k,p) \\
    \end{bmatrix}
\end{equation}
and
\begin{equation}\label{eq:11:2}
    \begin{bmatrix}
      S_{2RR}(k,p) \\
      S_{2LL}(k,p) \\
    \end{bmatrix}
    =
    \begin{bmatrix}
      e^{iEl} & 0 \\
      0 & e^{-iEl} \\
    \end{bmatrix}
    \begin{bmatrix}
      S_{1RR}(k,p) \\
      S_{1LL}(k,p) \\
    \end{bmatrix}.
\end{equation}

We combine the scattering matrix of the cavity, the free space and the emitter in sequence to get the scattering matrix of the whole system
\begin{equation}\label{eq:12}
  T_s =
  \begin{bmatrix}
      t^2-\frac{r^4}{t^2} & \frac{r^2}{t^2} \\
      -\frac{r^2}{t^2} & \frac{1}{t^2} \\
  \end{bmatrix}
  \begin{bmatrix}
      e^{iEl} & 0 \\
      0 & e^{-iEl} \\
  \end{bmatrix}
  \begin{bmatrix}
      \hat{T}-\hat{R}\hat{T}^{-1}\hat{R} & \hat{R}\hat{T}^{-1} \\
      -\hat{T}^{-1}\hat{R} & \hat{T}^{-1} \\
  \end{bmatrix}
  \begin{bmatrix}
      e^{iEl} & 0 \\
      0 & e^{-iEl} \\
  \end{bmatrix}
  \begin{bmatrix}
      t^2-\frac{r^4}{t^2} & \frac{r^2}{t^2} \\
      -\frac{r^2}{t^2} & \frac{1}{t^2} \\
    \end{bmatrix}.
\end{equation}

Next, the procedure of our approach is described as follow [below the wave packets are shown in Fig.~\ref{fig:2}]
\begin{enumerate}
  \item By using the formulae in Tab.~\ref{tab:1} and Tab.~\ref{tab:2}, we derive the operators $\hat{T}_{s,22}^{-1}$ and $T_{s,12}T_{s,22}^{-1}$. So we get the frequency spectrums of wave packets \textcircled{\small{1}} and \textcircled{\small{4}}, which are $\hat{T}_{s,22}^{-1} f_{in}(k,p)$ and $T_{s,12}T_{s,22}^{-1} f_{in}(k,p)$ respectively \cite{FanAPL80}. Here $f_{in}(k,p)$ is the frequency spectrum of the incident two photons, which is a delta function and denoted by the uncircled number ``1'' in Fig.~\ref{fig:2}.
  \item By using the scattering matrix of the cavity [Eq.~(\ref{eq:11:1})], we derive the frequency spectrums of wave packets \textcircled{\small{2}} and \textcircled{\small{5}} from \textcircled{\small{1}} and wave packet \textcircled{\small{6}} from \textcircled{\small{4}}.
  \item By using the scattering matrix of the emitter [Eq.~(\ref{eq:1})], we derive the frequency spectrum of wave packet \textcircled{\small{7}} from \textcircled{\small{5}} and \textcircled{\small{6}}. When one photon of wave packet \textcircled{\small{7}} transmits the cavity, we get wave packet \textcircled{\small{3}} from \textcircled{\small{7}}.
  \item we calculate the two photon transmitted wave packets associated with \textcircled{\small{2}} and \textcircled{\small{3}}. In this situation, the problem turns to single photon transport, so we can use the results in Ref.~\cite{FanAPL80} directly.
  \item By adding the two photon transmitted wave packets associated with \textcircled{\small{2}} and \textcircled{\small{3}} to the wave packet \textcircled{\small{1}}, we derive the frequency spectrum of the two photon transmitted wave packet finally. After fourier transformation, we get the two-photon wave function $t_2(x_1,x_2)$ in part of the out states, in which both photons are transmitted.
\end{enumerate}

\section{Results}\label{sec:4}
\subsection{Situation for Low-Q Cavity}\label{sec:4:1}
Here we consider a low-Q cavity. We set the cavity reflectivity $r=0.9$, cavity transmission amplitude $t=i\sqrt{1-r^2}$, the coupling strength of the emitter $\Gamma=0.004(2\pi c/l)$.

\begin{figure}
  \centering
  \subfigure[Without cavity]{\label{fig:3a}
  \includegraphics[width=1.5in]{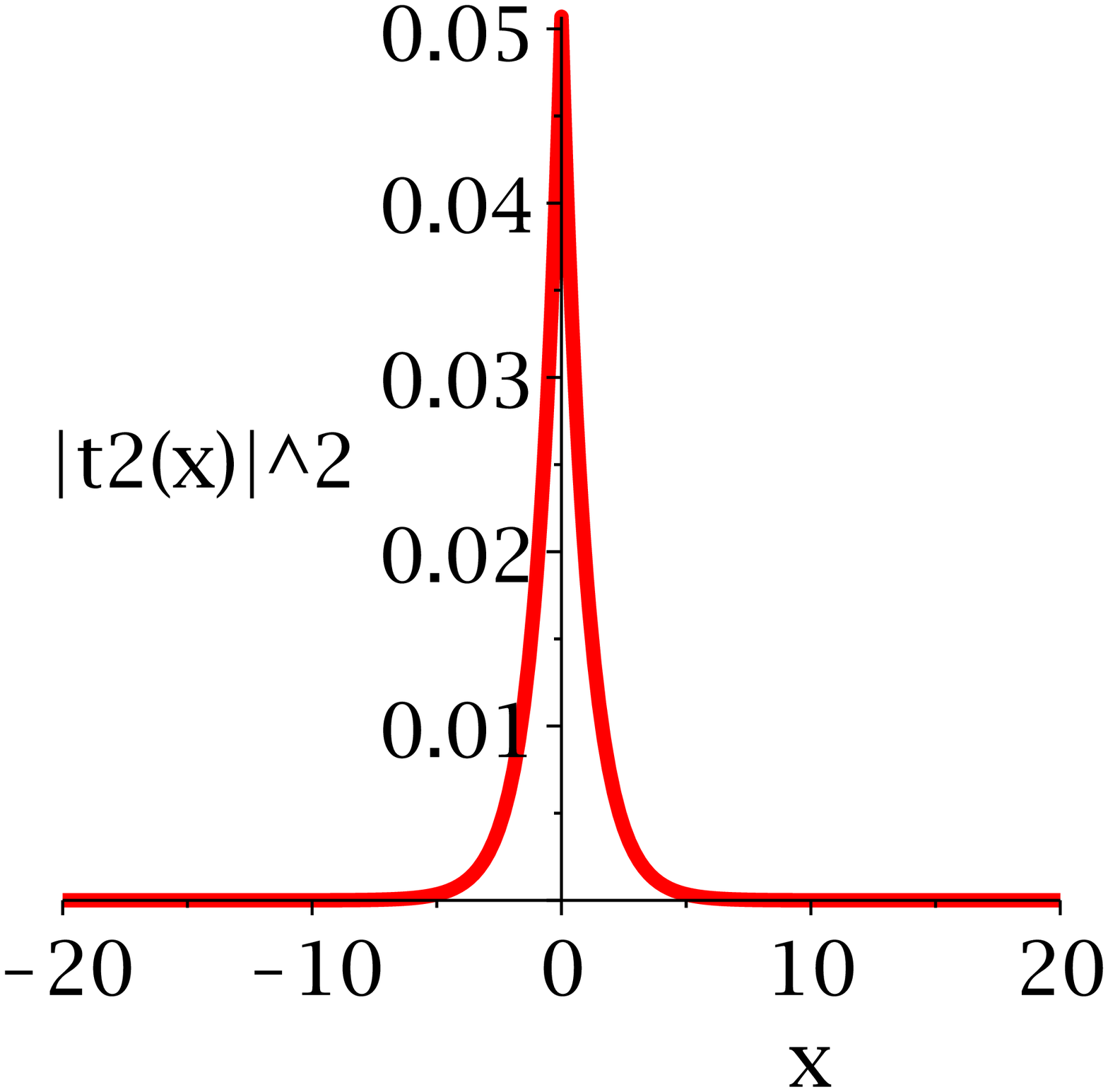}}\hspace{1em}
  \subfigure[$E_1/2=0.240(2\pi c/l)$]{\label{fig:3b}
  \includegraphics[width=1.5in]{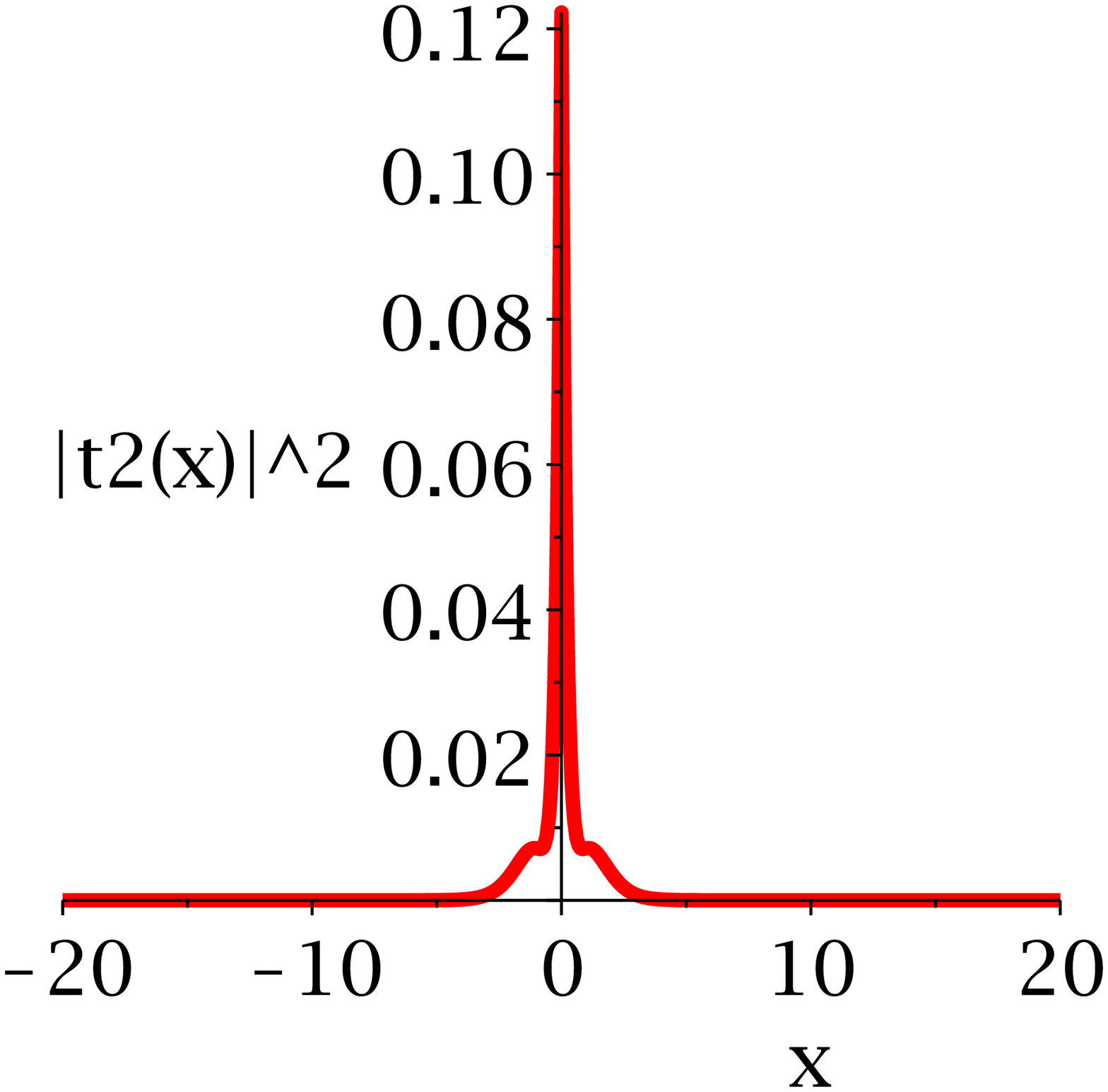}}\hspace{1em}
  \subfigure[$E_1/2=0.125(2\pi c/l)$]{\label{fig:3c}
  \includegraphics[width=1.5in]{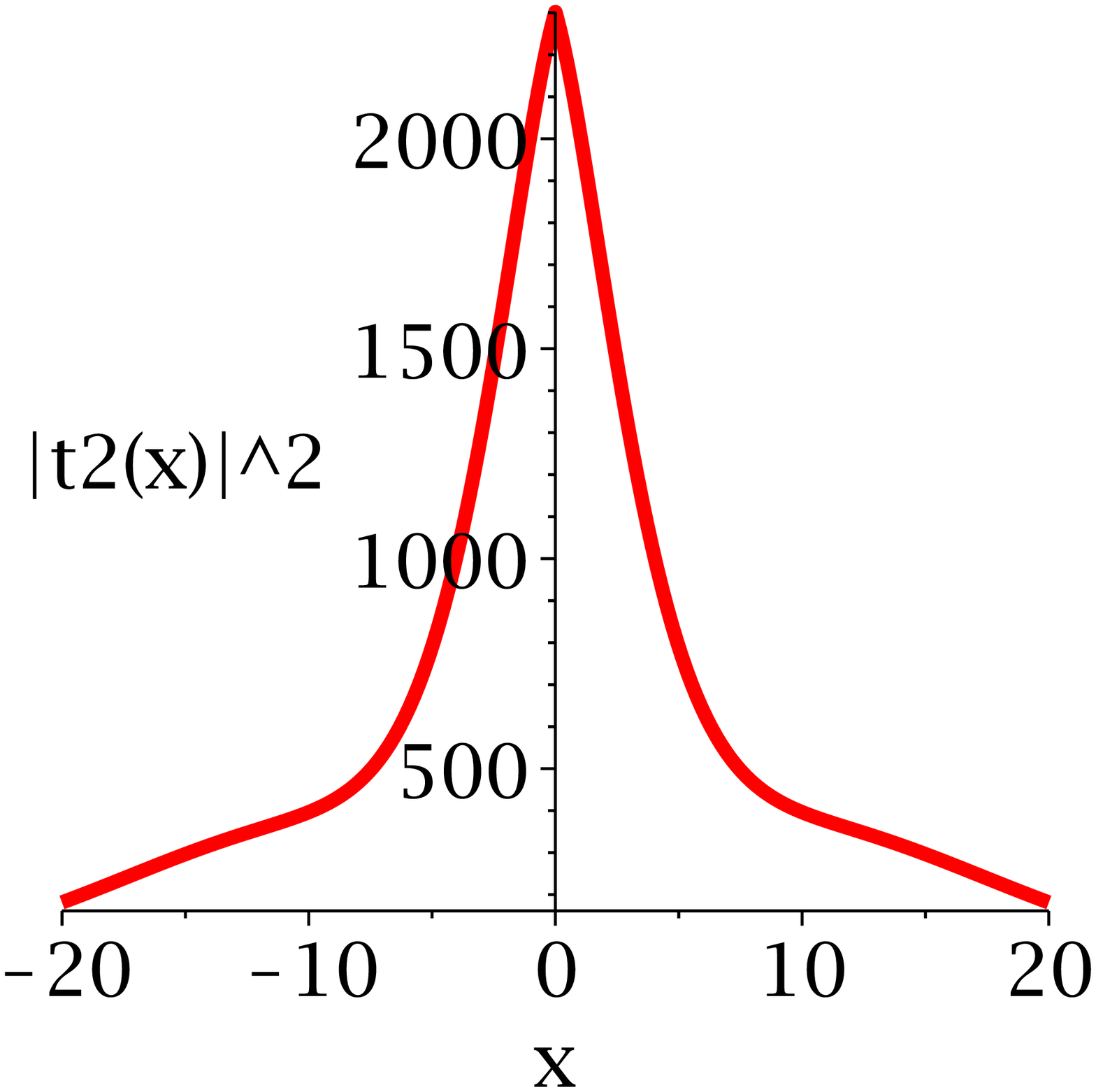}}\\
  \caption{Two-photon correlated transmission probability density $|t_2(x_1,x_2)|^2$. Two photons are near resonant with the emitter transition frequency $\Omega$. (a) shows the situation without the cavity. (b) and (c) represent the influences of the cavity. The abscissa is the distance difference $x=x_1-x_2$ of the two detectors ($D_1$, $D_2$) [Fig.~\ref{fig:1a}], which is scaled by $1/\Gamma$.}\label{fig:3}
\end{figure}

Firstly, We consider the situation that two photons are near resonant with the emitter transition frequency $\Omega$ [i.e. $\delta E \simeq 0, \Delta \simeq 0$, see Fig.~\ref{fig:1b}]. We find that, in ordinary case [e.g. $E_1/2=0.240(2\pi c/l)$, see Fig.~\ref{fig:3b}], the correlated transmission probability density $|t_2(x_1,x_2)|^2$ contracts, which indicates that the transmitted light has a wider frequency spectrum than the situation without the cavity \cite{FanPRL98, FanPRA76}. This is because the photons reflected and scattered many times. But when the two photons are resonant with the cavity resonance reflection frequency [e.g. $E_1/2=0.125(2\pi c/l)$, see Fig.~\ref{fig:3c}], the two-photon wave packet extends. So the cavity makes the frequency spectrum of the transmitted light narrower than that without the cavity. As a consequence, we achieve a strong photon-photon interaction by properly tuning the resonant frequency of the cavity, which is important for realization of controlled optical nonlinearity.

\begin{figure}
  \centering
  \subfigure[$E/2=0.322(2\pi c)/l$]{\label{fig:4a}
    \includegraphics[width=1.5in]{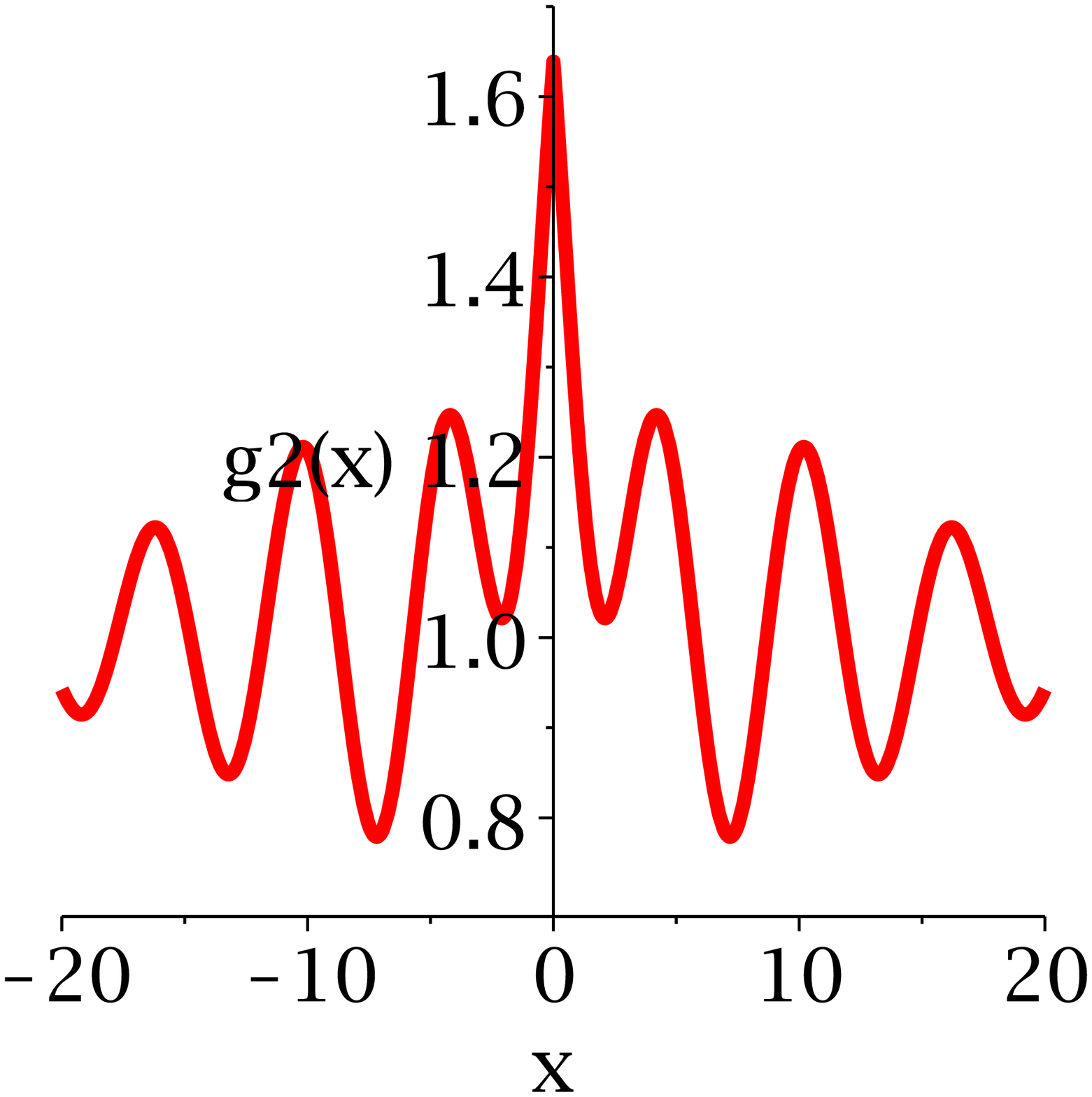}}\hspace{1em}
  \subfigure[$E/2=0.321(2\pi c)/l$]{\label{fig:4b}
    \includegraphics[width=1.5in]{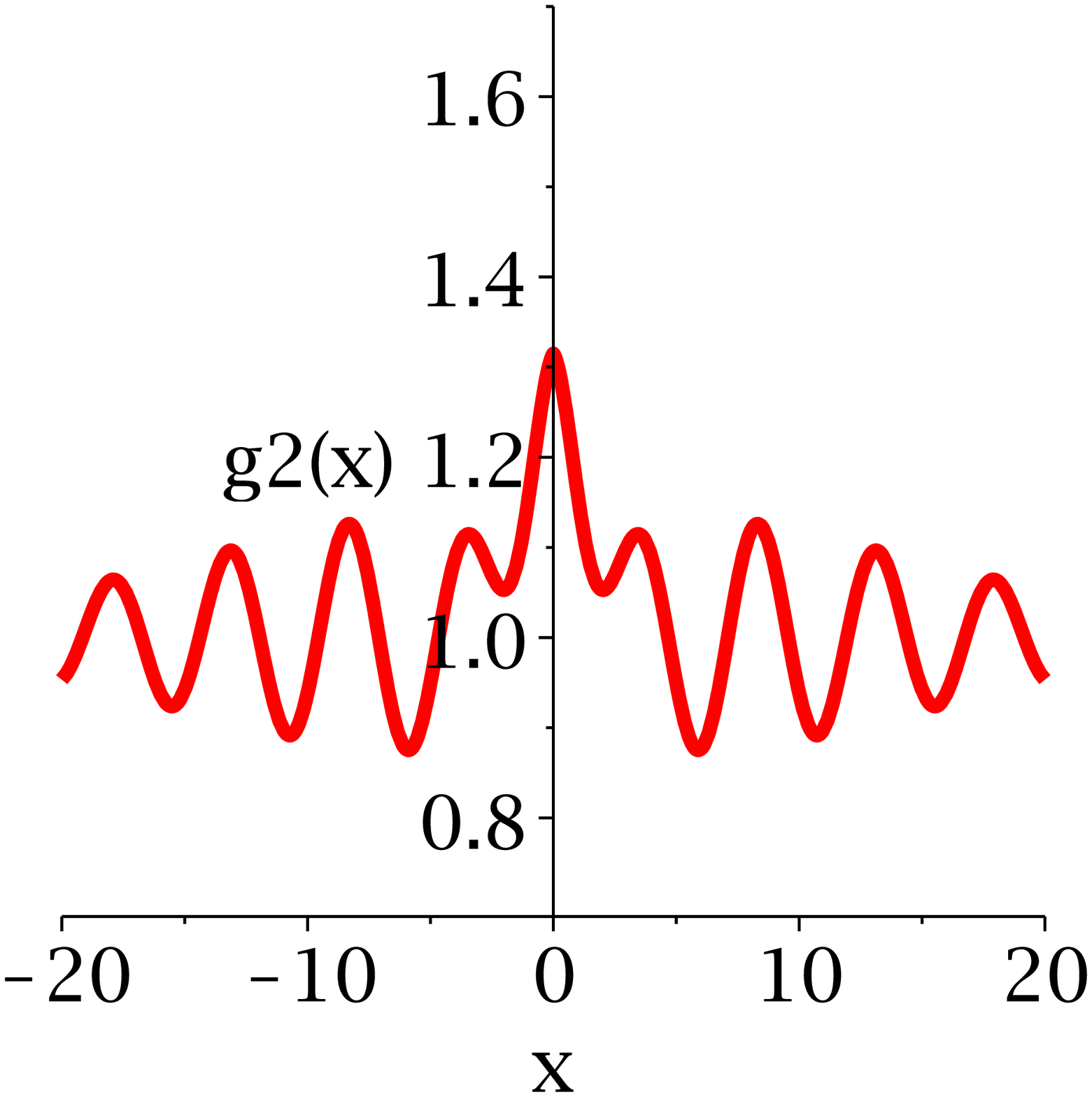}}\hspace{1em}
  \subfigure[$E/2=0.320(2\pi c)/l$]{\label{fig:4c}
    \includegraphics[width=1.5in]{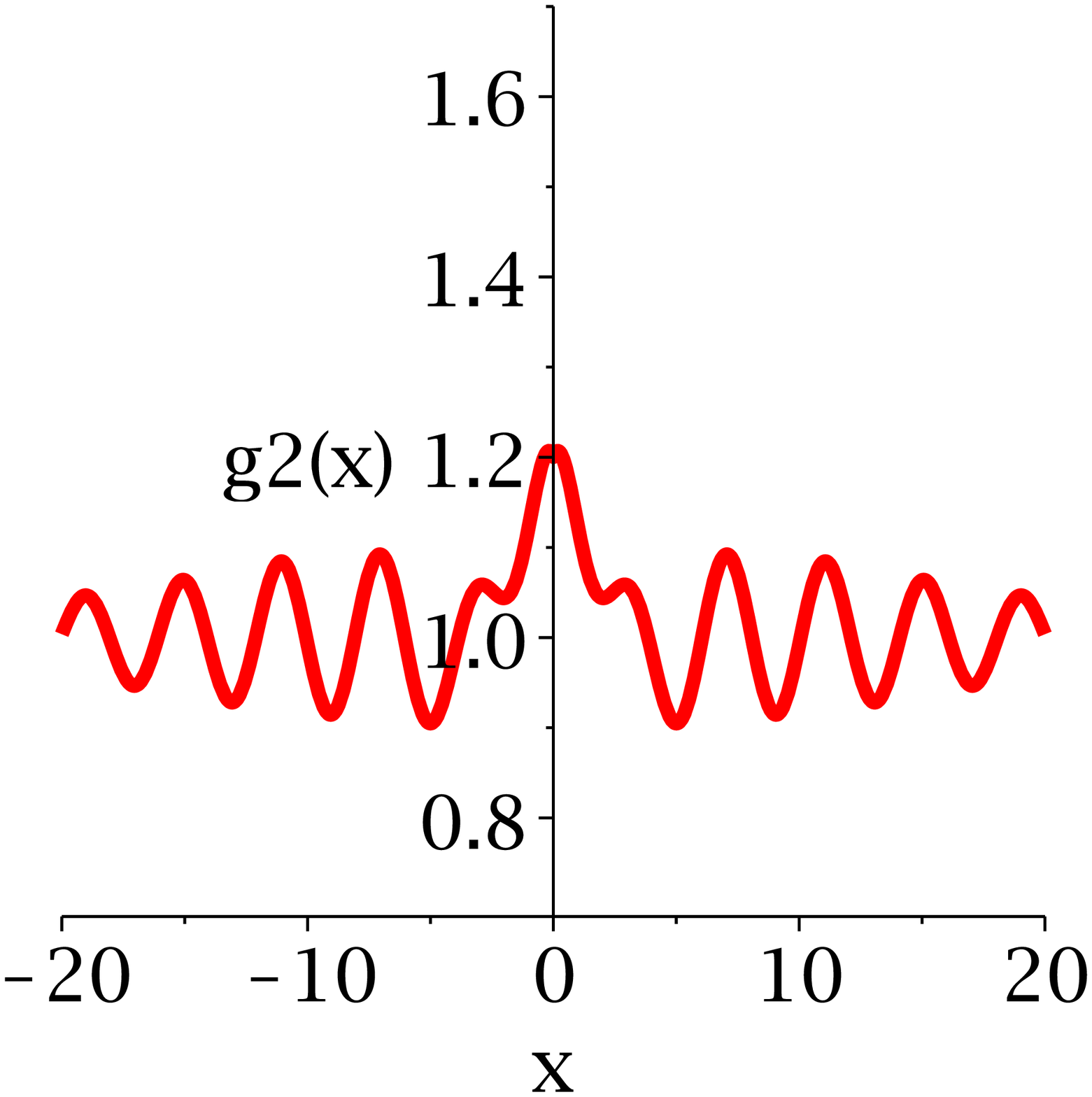}}\hspace{1em}
  \subfigure[$E/2=0.328(2\pi c)/l$]{\label{fig:4d}
    \includegraphics[width=1.5in]{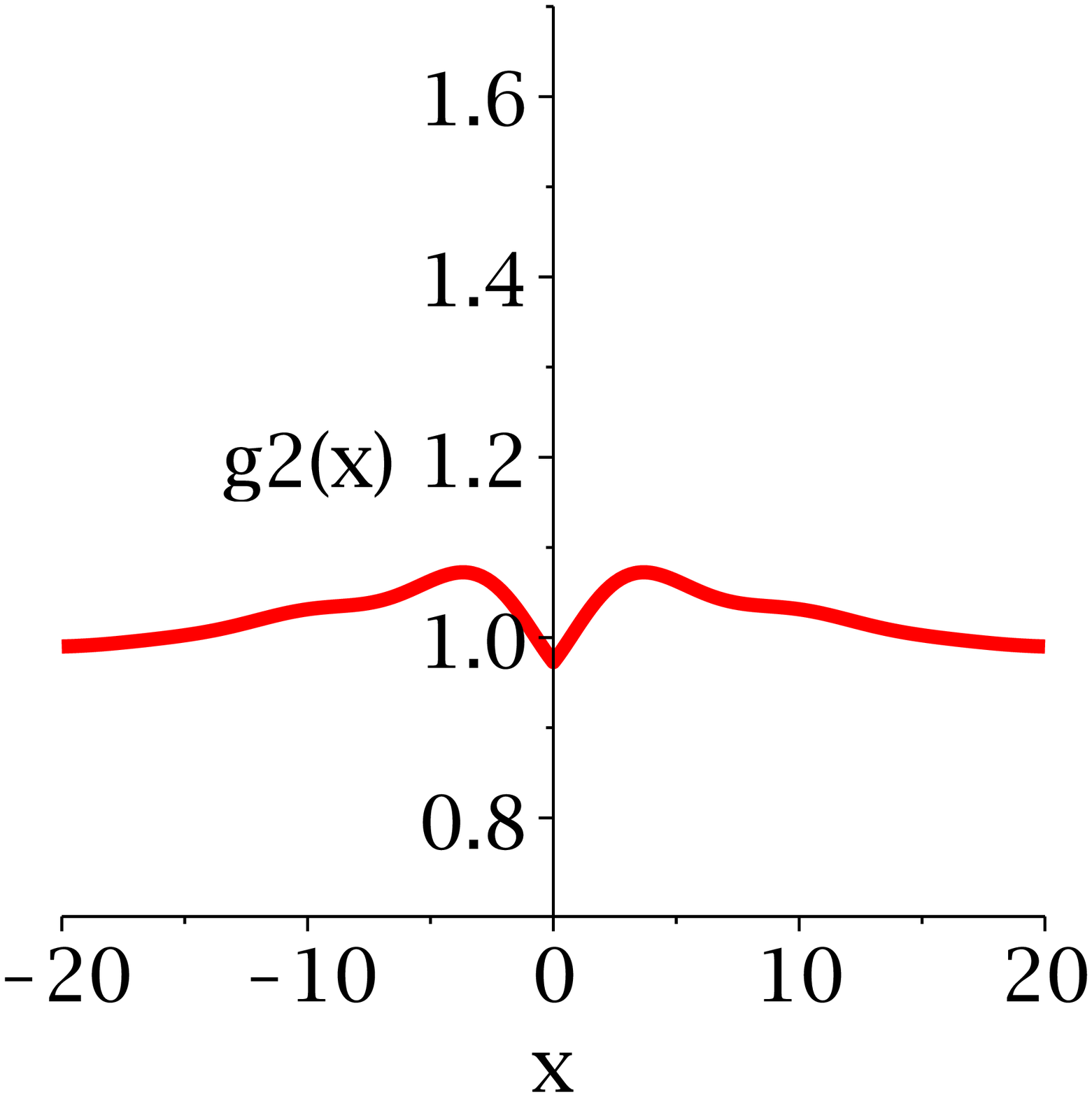}}\hspace{1em}
  \subfigure[$E/2=0.329(2\pi c)/l$]{\label{fig:4e}
    \includegraphics[width=1.5in]{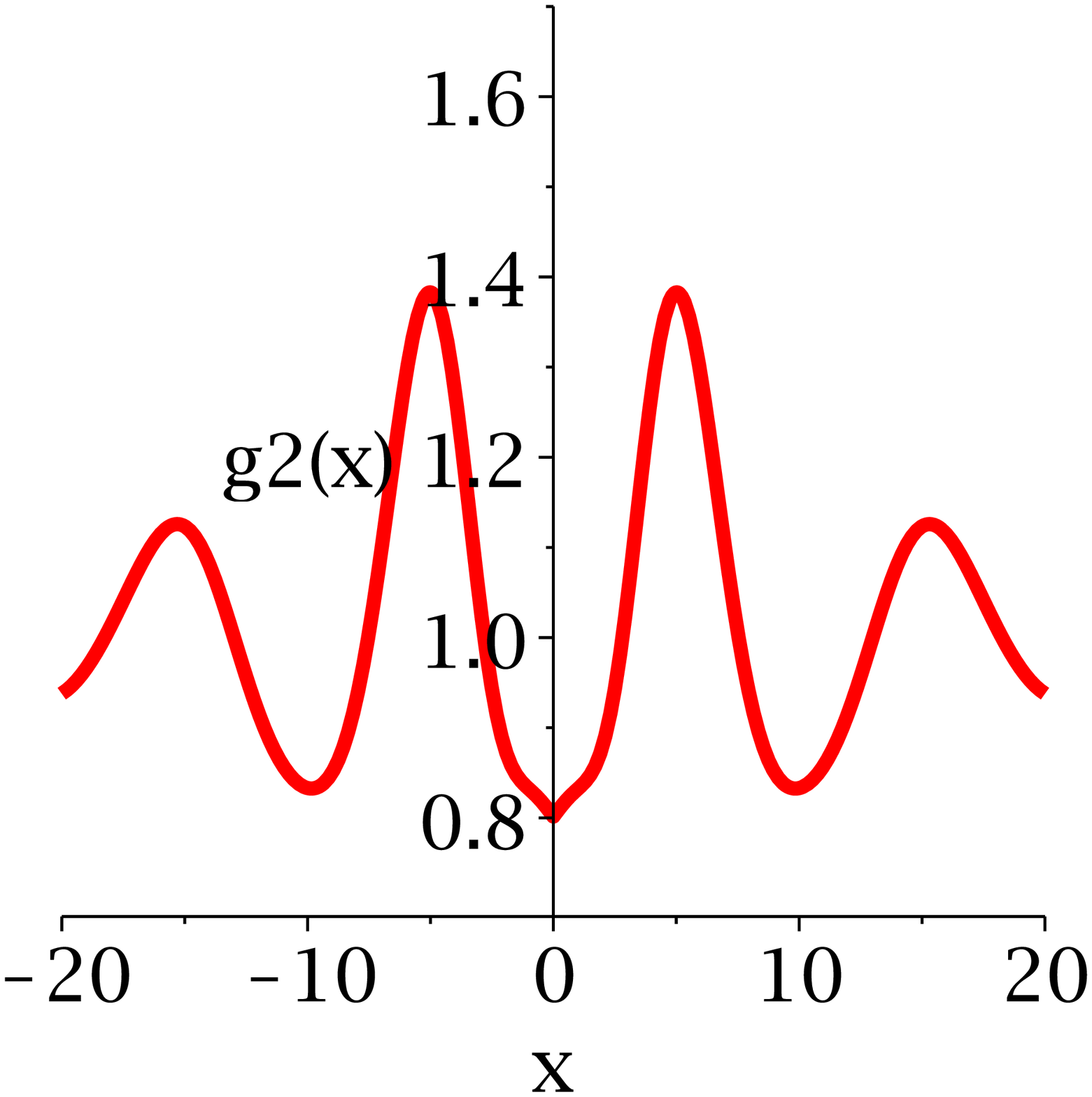}}\hspace{1em}
  \subfigure[$E/2=0.330(2\pi c)/l$]{\label{fig:4f}
    \includegraphics[width=1.5in]{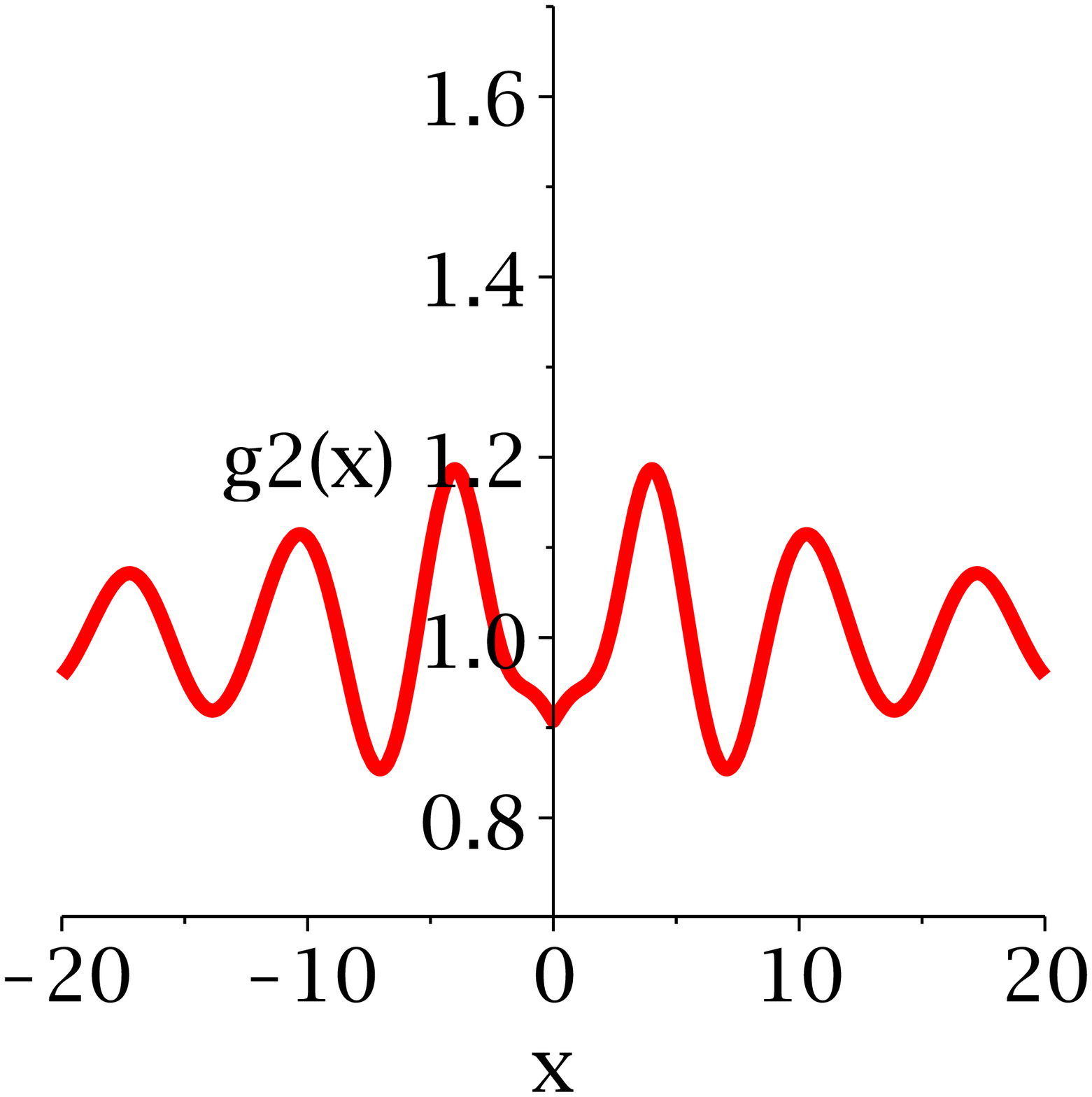}}\\
  \caption{Intensity correlation function $g^{(2)}(x_1,x_2)$. Two photons are non-resonant with the emitter transition frequency $\Omega=0.325(2\pi c)/l$. The incident two-photon momenta difference $\Delta=0$ [see Fig.~\ref{fig:1b}]. The abscissa is the same as that in Fig.~\ref{fig:3}.}\label{fig:4}
\end{figure}

Then we study the non-resonant case that two photons are not resonant with the emitter transition frequency $\Omega=0.325(2\pi c)/l$ (Fig.~\ref{fig:4}). Here we consider the intensity correlation function $g^{(2)}(x_1,x_2)$, which is defined as follow \cite{ScullyQO}
\begin{equation}\label{eq:13}
    g^{(2)}(x_1,x_2) = g^{(2)}(x) = \frac {\< a^+(x_1)a^+(x_2)a(x_2)a(x_1) \>} {\< a^+a \>^2} = \frac {|t_2(x_1,x_2)|^2} {D},
\end{equation}
where $x=x_1-x_2$. $D$ is the normalization constant and independent of $x$.

We show that the intensity correlation functions $g^{(2)}(x_1,x_2)$ are nearly opposite to each other at the two sides of the emitter transition frequency $\Omega$ rather than be the same. The transmitted two photons exhibit bunching behaviors and super-Poissonian statistics ($g^{(2)}(0)>1$) when $E/2<\Omega$ (the upside figures of Fig.~\ref{fig:4}) and anti-bunching behaviors and sub-Poissonian statistics ($g^{(2)}(0)<1$) when $E/2<\Omega$ (the underside figures of Fig.~\ref{fig:4}). We already know that, for single photons, Fano resonance exhibits a sharp asymmetric line shape with the transmission coefficients varying from 0 to 100\% over a very narrow frequency range \cite{FanJOSAA20}. So here we show the Fano resonance line shape for two photons. Because of its sharp asymmetric line shape, such an effect is important for lowering the power threshold in optical bistable devices and for sensing applications \cite{FanAPL80}.

\subsection{Situation for High-Q Cavity}\label{sec:4:2}
\begin{figure}
  \centering
  \subfigure[$log(g^{(2)}(0))-E/2$]{\label{fig:0a}
  \includegraphics[width=1.5in]{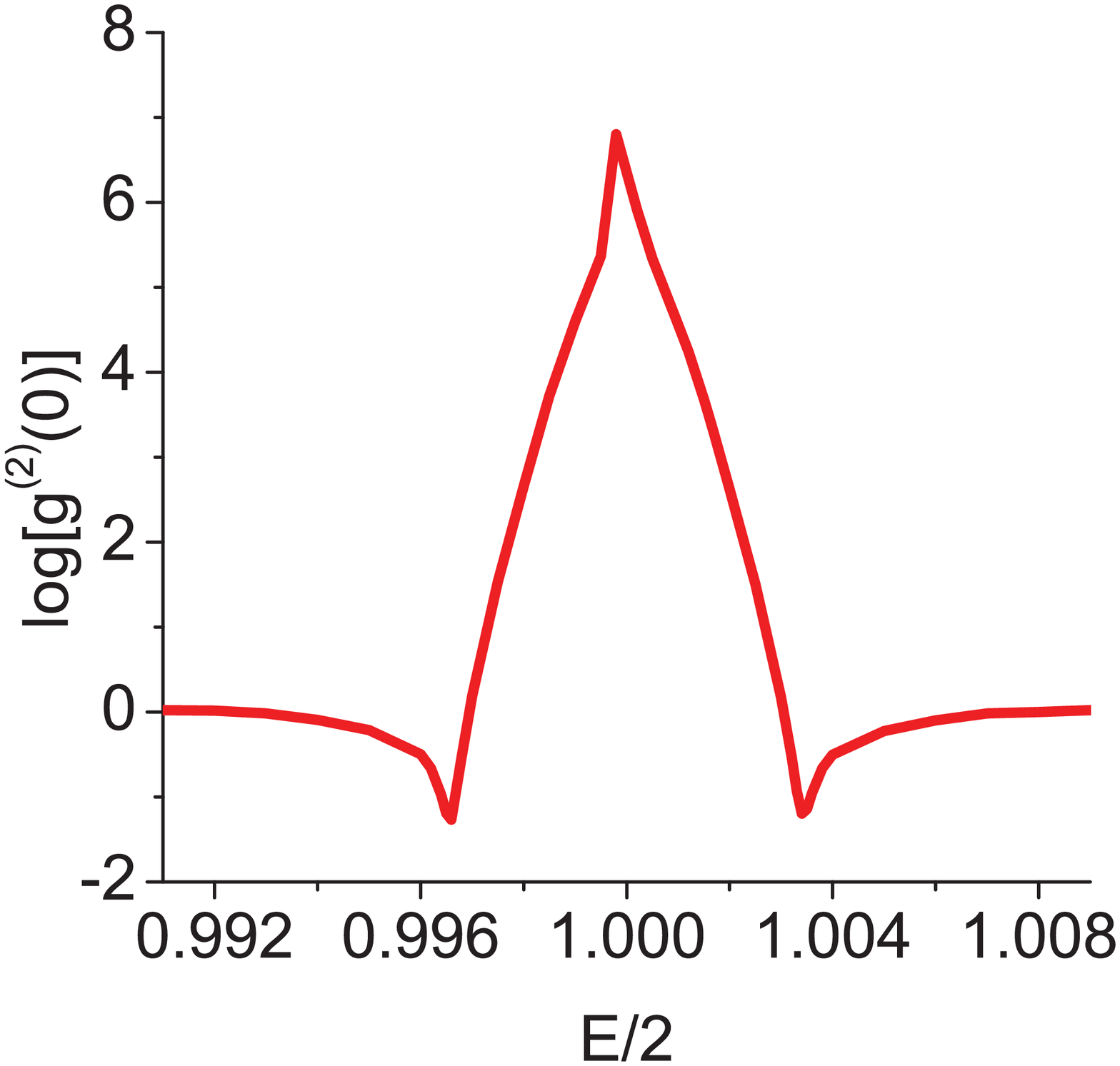}}\hspace{1em}
  \subfigure[$g^{(2)}(x)-x$]{\label{fig:0b}
  \includegraphics[width=1.5in]{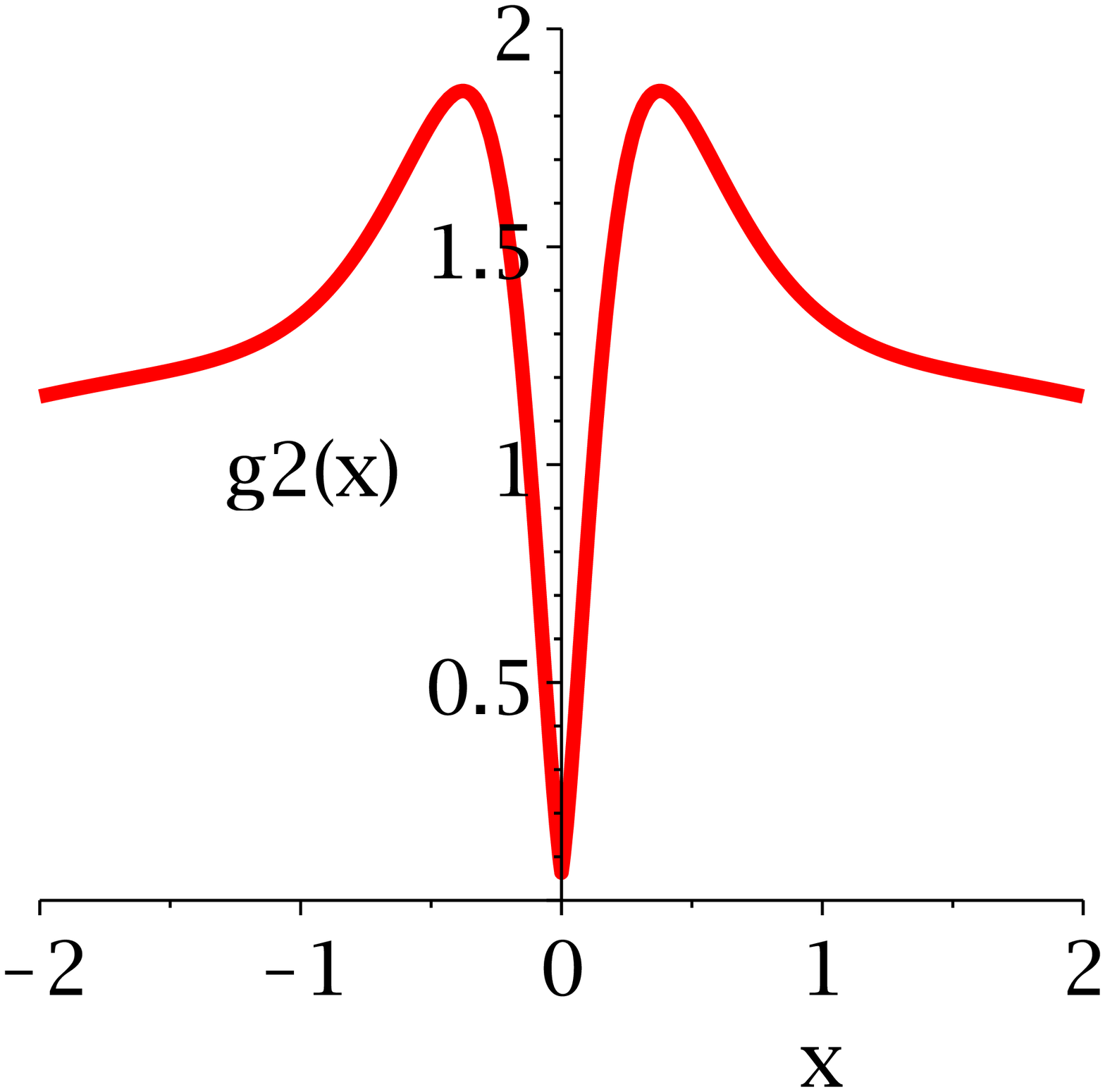}}\hspace{1em}
  \subfigure[$g^{(2)}(x)-x$]{\label{fig:0c}
  \includegraphics[width=1.5in]{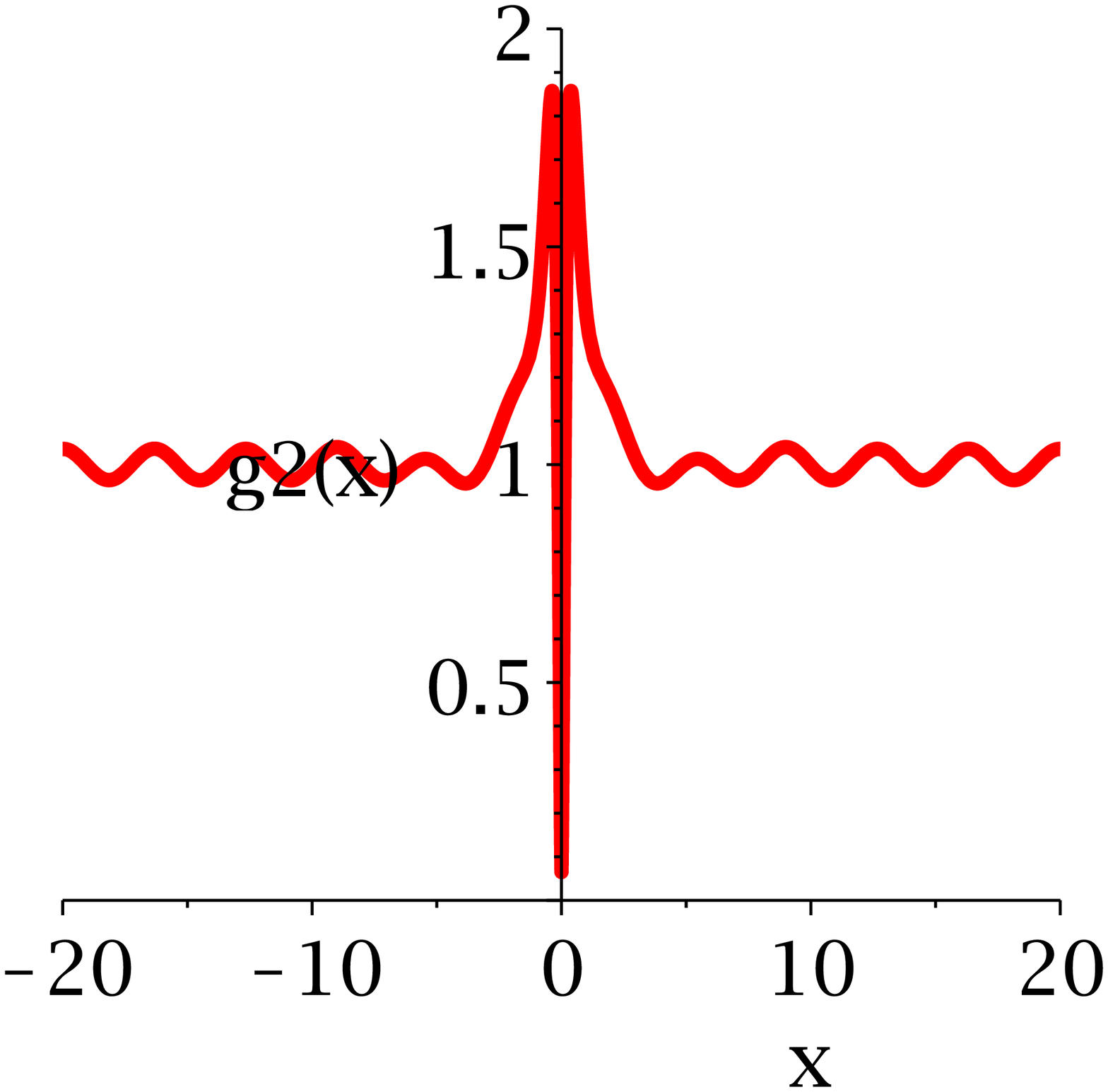}}\\
  \caption{Intensity correlation function $g^{(2)}(x)$ for a high-Q cavity. The abscissa of (a) is scaled by $(2\pi c/l)$. The abscissae of (b) and (c) are the same as that in Fig.~\ref{fig:3}. Here (b) is different from (c) in the range of the abscissa.}\label{fig:0}
\end{figure}

In a high-Q cavity, we can compare our results with the recent works done by Birnbaum et al. \cite{BirnbaumNat436} and Shi et al. \cite{ShiarXiv1009}. We set our parameters as follow to fit their conditions: the cavity's reflectivity $r=0.9996$ and transmission amplitude $t=i\sqrt{1-r^2}$, the emitter transition frequency $\Omega=2\pi c/l$ (which is also the cavity resonance frequency when $r$ is a real number), the coupling strength $\Gamma=0.002(2\pi c/l)$. We also consider the intensity correlation function $g^{(2)}(x_1,x_2) = g^{(2)}(x)$ (Fig.~\ref{fig:0}).

Fig.~\ref{fig:0a} represents the relation between $log[g^{(2)}(0)]$ and $E/2$, which agrees with the results in Ref.~\cite{BirnbaumNat436, ShiarXiv1009}. When $E/2=0.9966(2\pi c/l)$ or $1.034(2\pi c/l)$, $log[g^{(2)}(0)]$ gets its minimum. Comparing with Ref.~\cite{BirnbaumNat436, ShiarXiv1009}, we also consider the situation when $E/2=0.9966(2\pi c/l)$ [Fig.~\ref{fig:0b} and Fig.~\ref{fig:0c}]. Because our model includes the decay of the cavity inherently, our results exactly match the Figure.~3 of Ref.~\cite{BirnbaumNat436}, except where has a modulation arising from center-of-mass motion of the trapped atom. In Ref.~\cite{ShiarXiv1009}, however, the $g^{(2)}(\tau)$ arises to unity after infinite time because it does not consider the decay of the cavity or that of the two-level emitter.

\section{Summary}\label{sec:5}
In conclusion, we study the two-photon correlated scattering by a two-level cavity-coupled emitter in a 1D waveguide. Our results agree with the recent works \cite{BirnbaumNat436, ShiarXiv1009}. Our research is important for realization of strong photon-photon interaction at the case of few photons and for quantum network. In addition, we propose a general approach to two-photon, cavity-coupled and multi-frequency scattering problem in the 1D waveguide. The same approach can deal with the three-level emitter, which is useful to make the photon-transistor \cite{ChangNatphy3, WittNJP12}.

\section{Acknowledgments}
We acknowledge the financial support of Special Prophase Project on the National Basic Research Program of China (No.2011CB311807), the National Natural Science Foundation of China (No.10804092), the Natural Science Basic Research Plan in Shaanxi Province of China (No.2010JQ1004), the Fundamental Research Funds for the Central Universities (No.xjj20100099).

\appendix

\section{Derivations of Operator-Multiplying Formula}\label{app:1}
Here we derive the Operator-multiplying formula in Tab.~\ref{tab:1}. If operators $\hat{T}$ and $\hat{R}$ are defined as follow
\begin{equation}\label{app:eq:1}
\begin{gathered}
    g(k_2,p_2) = \hat{T} f(k,p) = \frac{1}{2} \int_{0}^{\infty} dk_1 dp_1 f(k_1,p_1) \times \\
    \{ tt(k_1,p_1) [ \delta(k_{1}-k_2) \delta(p_{1}-p_2) + \delta(k_{1}-p_2) \delta(p_{1}-k_2) ] + B(\Delta_1,\Delta_2) \delta(E_{1}-E_2) \},
\end{gathered}
\end{equation}
\begin{equation}\label{app:eq:2}
\begin{gathered}
    \hat{R} g(k_2,p_2) = \frac{1}{2} \int_{0}^{\infty} dk_2 dp_2 g(k_2,p_2) \times \\
    \{ rr(k_2,p_2) [ \delta(k_{2}-k) \delta(p_{2}-p) + \delta(k_{2}-p) \delta(p_{2}-k) ] + C(\Delta_2,\Delta) \delta(E_{2}-E) \},
\end{gathered}
\end{equation}
then the operator $\hat{R}\hat{T}$ can be derived as follow
\begin{equation*}
\begin{aligned}
    \hat{R}\hat{T} f(k,p) & = && \frac{1}{4} \int_{0}^{\infty} \int_{0}^{\infty} dk_1 dp_1dk_2 dp_2 f(k_1,p_1) \times \\
    & && \{ tt(k_1,p_1) rr(k_2,p_2) [ \delta(k_{1}-k_2) \delta(p_{1}-p_2) + \delta(k_{1}-p_2) \delta(p_{1}-k_2) ] \times \\
    & && [ \delta(k_{2}-k) \delta(p_{2}-p) + \delta(k_{2}-p) \delta(p_{2}-k) ] \\
    & && + B(\Delta_1,\Delta_2) rr(k_2,p_2) [ \delta(k_{2}-k) \delta(p_{2}-p) \\
    & && + \delta(k_{2}-p) \delta(p_{2}-k) ] \delta(E_{1}-E_2) \\
    & && + C(\Delta_2,\Delta) tt(k_1,p_1) [ \delta(k_{1}-k_2) \delta(p_{1}-p_2) \\
    & && + \delta(k_{1}-p_2) \delta(p_{1}-k_2)] \delta(E_{2}-E) \\
    & && + B(\Delta_1,\Delta_2) C(\Delta_2,\Delta) \delta(E_{1}-E_2) \delta(E_{2}-E) \}
\end{aligned}
\end{equation*}
\begin{equation}\label{app:eq:3}
\begin{aligned}
    & = && \frac{1}{2} \int_{0}^{\infty} dk_1dp_1 f(k_1,p_1) \times \\
    & && \{ tt(k_1,p_1) rr(k_1,p_1) [ \delta(k_{1}-k) \delta(p_{1}-p) + \delta(k_{1}-p) \delta(p_{1}-k) ] \\
    & && + [B(\Delta_1,\Delta) rr(k,p) + C(\Delta_1,\Delta) rr(k_1,p_1) \\
    & && + \int_{-\infty}^{\infty} B(\Delta_1,\Delta_2) C(\Delta_2,\Delta) \delta(E_{2}-E_{1}) d\Delta_2]\delta(E_{1}-E) \} \\
    & = && \frac{1}{2} \int_{0}^{\infty} dk_1dp_1 f(k_1,p_1) \times \\
    & && \{ tt(k_1,p_1) rr(k_1,p_1) [ \delta(k_{1}-k) \delta(p_{1}-p) + \delta(k_{1}-p) \delta(p_{1}-k) ] \\
    & && +[B_0B_1(\Delta_1)B_2(\Delta_2) rr(\Delta) + C_0C_1(\Delta_1)C_2(\Delta_2) tt(\Delta_1) \\
    & && + B_0C_0B_1(\Delta_1)C_2(\Delta_2)f_{int}]\delta(E_{1}-E) \}.
\end{aligned}
\end{equation}
Here we factorize the background fluorescence $B(\Delta_1,\Delta_2)$ and $C(\Delta_1,\Delta_2)$ as follow
\begin{equation}\label{app:eq:4}
\begin{aligned}
    B(\Delta_1,\Delta_2) & = B_0B_1(\Delta_1)B_2(\Delta_2), \\
    C(\Delta_1,\Delta_2) & = C_0C_1(\Delta_1)C_2(\Delta_2).
\end{aligned}
\end{equation}
For example, if $\hat{T}=\hat{T}_{22}$ [Eq.~(\ref{eq:2})], then
\begin{equation*}
\begin{gathered}
    B(\Delta_1,\Delta_2) = \frac {4i \Gamma^2} {\pi} \frac {E_1 - 2 \Omega + i\Gamma} {[4\Delta_1^2 - (E_1 - 2\Omega + i\Gamma)^2] [4\Delta^2 - (E_1 - 2\Omega + i\Gamma)^2]}, \\
    B_0 = \frac {4i \Gamma^2} {\pi} (E_1 - 2 \Omega + i\Gamma), \\
    B_1(\Delta_1) = \frac{1}{4\Delta_1^2 - (E_1 - 2\Omega + i\Gamma)^2}, \\
    B_2(\Delta_2) = \frac{1}{4\Delta_2^2 - (E_1 - 2\Omega + i\Gamma)^2}.
\end{gathered}
\end{equation*}
In Eq.~(\ref{app:eq:3})
\begin{equation}\label{app:eq:5}
    f_{int} = \frac{1}{2} \int_{-\infty}^{+\infty} B_2(y) C_1(y) dy.
\end{equation}

In Eq.~(\ref{app:eq:1}), we find that an operator can be determined by the part multiplied by $[ \delta(k_{1}-k) \delta(p_{1}-p) + \delta(k_{1}-p) \delta(p_{1}-k) ]$ (e.g. $tt(k_1,p_1)$ in operator $\hat{T}$) and the part multiplied by $\delta(E_{1}-E_2)$ (e.g. $B(\Delta_1,\Delta_2)$ in operator $\hat{T}$). So we briefly write the result of Eq.~(\ref{app:eq:3}) in Tab.~\ref{tab:1}.

\section{Derivations of Operator-inversion Formula}\label{app:2}
We use the operator-multiplying formula to derive the operator-inversion formula in Tab.~\ref{tab:2}. We order that $\hat{R}=\hat{T}^{-1}$, so we have
\begin{equation}\label{app:eq:6}
\begin{gathered}
  \hat{R} \hat{T} f(k,p) = \hat{T} \hat{R} f(k,p) = f(k,p) = \frac{1}{2} [f(k,p)+f(p,k)] \\
    = \frac{1}{2} \int dk_1 dp_1 f(k_1,p_1) [ \delta(k_{1}-k) \delta(p_{1}-p) + \delta(k_{1}-p) \delta(p_{1}-k) ].
\end{gathered}
\end{equation}
Comparing Eq.~(\ref{app:eq:6}) with Eq.~(\ref{app:eq:3}), we get
\begin{equation}\label{app:eq:7}
\begin{gathered}
    tt(k_1,p_1) rr(k_1,p_1) = 1, \\
    B_0B_1(\Delta_1)B_2(\Delta_2) rr(\Delta) + C_0C_1(\Delta_1)C_2(\Delta_2) tt(\Delta_1) + B_0C_0B_1(\Delta_1)C_2(\Delta_2)f_{int} = 0.
\end{gathered}
\end{equation}
So
\begin{equation}\label{app:eq:7}
\begin{gathered}
    C_0 = -\frac{B_0}{1+B_0f_{int}}, \\
    C_1(\Delta_1) = \frac{B_1(\Delta_1)}{tt(\Delta_1)}, \\
    C_2(\Delta_2) = \frac{B_2(\Delta_2)}{tt(\Delta_2)}.
\end{gathered}
\end{equation}
Here
\begin{equation}\label{app:eq:8}
    f_{int} = \frac{1}{2} \int_{-\infty}^{+\infty} B_2(y) C_1(y) dy = \frac{1}{2} \int_{-\infty}^{+\infty} \frac{B_1(y) B_2(y)}{tt(y)} dy \equiv g_{int}.
\end{equation}

We briefly write the result of Eq.~(\ref{app:eq:7}) in Tab.~\ref{tab:2}.

\section{Derivations of the Reduced Scattering Matrix of the Emitter}\label{app:3}
Bellow we derive Eq.~(\ref{eq:9}) and Eq.~(\ref{eq:10}). From the third equation of Eq.~(\ref{eq:1}) we have
\begin{equation}\label{app:eq:9}
    S_{RLout}(k,p) = \hat{T}_{21}S_{1RR}(k,p) + \hat{T}_{21}S_{2LL}(k,p) + \hat{T}_{21}S_{RLin}(k,p).
\end{equation}
By using the following relation (see Fig.~\ref{fig:2})
\begin{equation}\label{app:eq:10}
    S_{RLin}(k,p) = r'^2 S_{RLout}(k,p),
\end{equation}
we have
\begin{equation}\label{app:eq:11}
    S_{RLin}(k,p) = r'^2 (1-r'^2\hat{R}_{11})^{-1} \hat{T}_{21} [S_{1RR}(k,p)+S_{2LL}(k,p)].
\end{equation}
By putting Eq.~(\ref{app:eq:11}) into the first and second equations of Eq.~(\ref{eq:1})
\begin{equation}\label{app:eq:12}
\begin{aligned}
    S_{1LL}(k,p) &= \hat{R}_{22}S_{1RR}(k,p) + \hat{T}_{22}S_{2LL}(k,p) + \hat{T}_{12}S_{RLin}(k,p), \\
    S_{2RR}(k,p) &= \hat{T}_{22}S_{1RR}(k,p) + \hat{R}_{22}S_{2LL}(k,p) + \hat{T}_{12}S_{RLin}(k,p),
\end{aligned}
\end{equation}
we get
\begin{equation}\label{app:eq:13}
\begin{aligned}
    \hat{T} &= \hat{T}_{22} + r'^2 \hat{T}_{12} (1-r'^2\hat{R}_{11})^{-1} \hat{T}_{21}, \\
    \hat{R} &= \hat{R}_{22} + r'^2 \hat{T}_{12} (1-r'^2\hat{R}_{11})^{-1} \hat{T}_{21}.
\end{aligned}
\end{equation}
Here we make the substitution
\begin{equation}\label{eq:app:14}
    r'^2 = \sqrt{2}e^{2iEl} \times r^2,
\end{equation}
where $r$ is the reflectivity of the cavity.

By using the formulae in Tab.~\ref{tab:1} and Tab.~\ref{tab:2}, we get the expressions of $\hat{T}$ and $\hat{R}$.










\end{document}